\documentclass[aps,prl,twocolumn,superscriptaddress,showpacs]{revtex4-1}
\usepackage[dvips]{graphicx}
\usepackage{multirow,slashbox}
\DeclareTextSymbol{\degre}{T1}{6}
\DeclareTextSymbol{\degre}{OT1}{23}

\begin{document}

\title{Competing magnetic ground states in non-superconducting Ba(Fe$_{1-x}$Cr$_x$)$_2$As$_2$}

\author{K. Marty}
\author{A. D. Christianson}
\author{C. H. Wang}
\author{M. Matsuda}
\author{H. Cao}
\affiliation{Oak Ridge National Laboratory, Oak Ridge, Tennessee 37831 USA}

\author{L. H. VanBebber}
\affiliation{Department of Materials Science and Engineering, University of Tennessee, Knoxville, Tennessee 37996 USA}
\author{J. L. Zarestky}
\affiliation{Ames Laboratory, U.S. DOE and Department of Physics and Astronomy, Iowa State University, Ames, Iowa 50011 USA}
\affiliation{Oak Ridge National Laboratory, Oak Ridge, Tennessee 37831 USA}
\author{D. J. Singh}
\author{A. S. Sefat}
\author{M. D. Lumsden}
\affiliation{Oak Ridge National Laboratory, Oak Ridge, Tennessee 37831 USA}

\date{\today}

\begin{abstract}

We present neutron diffraction measurements on single crystal samples of non-superconducting Ba(Fe$_{1-x}$Cr$_x$)$_2$As$_2$ as a function of Cr-doping for 0$\leq$x$\leq$0.47.
The average SDW moment is independent of concentration for x$\leq$0.2 and decreases rapidly for x$\geq$0.3.  For concentrations in excess of 30\% chromium, we find a new G-type antiferromagnetic phase which rapidly becomes the dominant magnetic ground state.  Strong magnetism is observed for all concentrations measured and competition between these ordered states and superconductivity naturally explains the absence of superconductivity in the Cr-doped materials.

\end{abstract}



\maketitle

The discovery of Fe-based superconductors \cite{Kamihara2008c} in multiple families of materials \cite{Rotter,Tapp,Hsu} with diverse doping possibilities enables systematic investigations which may ultimately lead to understanding of the essential physics underlying superconductivity. The interplay of magnetism and superconductivity\cite{Lumsden1,Mazin} has played a prominent role in the discussion based, partially, on the observation that superconductivity only emerges after sufficient suppression of the magnetically ordered state found in the parent compounds.

The parent compounds of the so-called 122 family of materials, AFe$_2$As$_2$ (A=Ba, Sr, Ca, and Eu), exhibit a structural phase transition, from tetragonal $I4/mmm$ to orthorhombic $Fmmm$ \cite{Lynn}, together with a simultaneous magnetic transition from a paramagnetic to antiferromagnetic spin-density-wave (SDW) state. Superconductivity only appears when these transitions are adequately suppressed. A key characteristic of the Fe-based materials is the variety of dopants which yield superconductivity.  In particular, superconductivity obtained by doping on the transition-metal site is a unique property of the iron pnictides, i.e., by partial substitution of Fe by Co \cite{Sefat1}, Ni \cite{Li}, Rh \cite{Ni}, Pd \cite{Ni}, Ir \cite{Han}, Pt \cite{Saha} or Ru \cite{Sharma,Thaler} in BaFe$_2$As$_2$.  From a chemical-doping perspective, the cases listed above are either isoelectronic substitution or electron-doping on the Fe site.  However, an exception occurs in the presence of hole doping;  replacement of Fe with either Cr \cite{Sefat2} or Mn \cite{Liu} causes suppression of the SDW and structural transitions but does not result in superconductivity.  The absence of superconductivity in these hole doped systems remains an unresolved issue.

To address this question, we have selected the Ba(Fe$_{1-x}$Cr$_x$)$_2$As$_2$ family of materials where  bulk measurements have previously determined the phase diagram for x up to 0.18\cite{Sefat2}. These studies showed that the magnetic and structural \cite{Budko} transitions are much more resilient to doping than, for example, Co-doped BaFe$_2$As$_2$ and, furthermore, did not yield superconductivity. The fact that these transitions are more robust in the case of Cr-doping does explain the lack of superconductivity. For example, doping with Ru causes the SDW state to persist to similar concentrations and, yet, superconductivity is still realized \cite{Sharma,Thaler}. In this letter, we present neutron diffraction measurements of Ba(Fe$_{1-x}$Cr$_x$)$_2$As$_2$ for 0$\leq$x$\leq$0.47.  The SDW and structural transition temperatures are indistinguishable for 0$\leq$x$\leq$0.335.  The ordered moment of the SDW state remains constant for most measured concentrations and decreases rapidly for x $\gtrsim$ 0.3.  As the moment is reduced, a competing phase, with an ordering wavevector consistent with G-type antiferromagnetic (AFM) order, appears.  First principles calculations suggest that the ground state energies of the SDW and G-type AFM states are comparable at x=0.25.  These results show that Cr-doping favors magnetism and that the suppression of the SDW phase is compensated by enhancement of a G-type AFM phase.  Strong magnetism across the entire phase diagram provides a natural explanation for the absence of superconductivity.



\begin{figure*}[t]
	\begin{center}
		\includegraphics[width=0.95\textwidth]{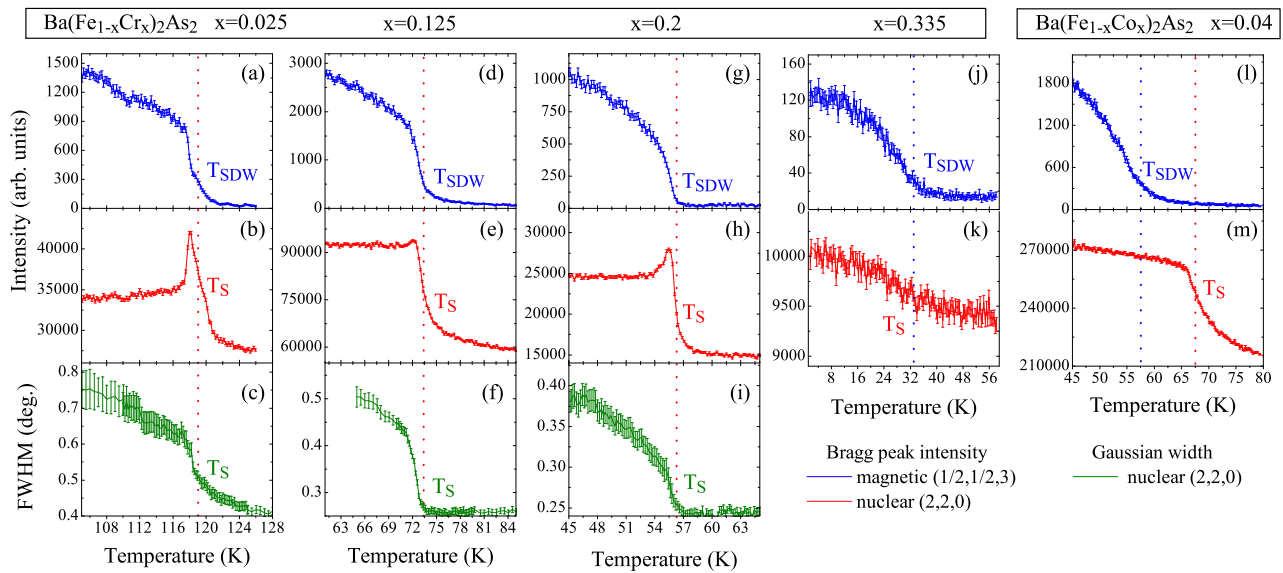}
		\caption{ (Color online) Ba(Fe$_{1-x}$Cr$_x$)$_2$As$_2$: x=0.025 ((a), (b), (c)), x=0.125 ((d), (e), (f)), x=0.2 ((g), (h), (i)), x=0.335 ((j), (k)) and Ba(Fe$_{1.96}$Co$_{0.04}$)$_2$As$_2$ ((l), (m)). Top row: intensity of the (1/2,1/2,3)$_T$ magnetic Bragg indicating the magnetic transition temperature T$_{SDW}$. Middle row: TAS measurements of the intensity of the nuclear Bragg peak (2,2,0)$_T$. The inflection point indicates the tetragonal to orthorhombic structural transition T$_S$. Bottom row: Gaussian full width at half maximum of the (2,2,0)$_T$ nuclear Bragg peak  measured on HB-3A, showing the structural transition T$_S$. Cr-x=0.335 and Co-x=0.04 were not measured on HB-3A.}
		\label{BigPanel}
	\end{center}
\end{figure*}

A mixture of FeAs and CrAs was used as a flux to synthesize the single crystals of Ba(Fe$_{1-x}$Cr$_x$)$_2$As$_2$\cite{Sefat2}. The chemical composition was determined using energy-dispersive x-ray spectroscopy from a JEOL JSM-840 scanning electron microscope\cite{Sefat2}. Eight different concentrations were studied : x = 0, 0.025, 0.075, 0.125, 0.2, 0.305, 0.335 and 0.47. The mass of the measured crystals range from 150mg to 4mg.

Neutron diffraction experiments were performed at the High Flux Isotope Reactor at Oak Ridge National Laboratory, using the HB-3A four-circle diffractometer, and the HB-1A, HB-1, and HB-3 triple-axis spectrometers (TAS).  Measurements on HB-3A used a wavelength of 1.56 \AA~with a flat monochromator to improve resolution.  The HB-1A measurements were performed with collimations of 48'-48'-40'-68' or 48'-open-open-open to increase flux for smaller samples.  HB-3 measurements were performed with 48'-open-open-open collimation.  Polarized neutron diffraction on HB-1 were performed with a focusing Heusler monochromator and flat Heusler analyzer with 48'-60'-40'-120' collimation and a small vertical guide field ($\sim$30 G) at the sample position yielding a flipping ratio of 15.3.

Figure \ref{BigPanel} shows representative, temperature dependent data for several Cr concentrations as well as data measured on Co-doped BaFe$_2$As$_2$ (x=0.04). The SDW transition temperature was determined from HB-1A measurements of the peak intensity of the (1/2,1/2,3)$_T$ magnetic reflection as shown in Fig. \ref{BigPanel} (a), (d), (g), (j), (l) (for clarity, we explicitly refer to wavevectors with a T or O subscript corresponding to tetragonal or orthorhombic basis).  The structural transition was studied using data from the HB-3A diffractometer and the HB-1A TAS.  Radial scans were performed on HB-3A to observe splitting of the tetragonal (2,2,0)$_T$ peak into the low temperature (4,0,0)$_O$ and (0,4,0,)$_O$ reflections.  The transition temperature was determined by plotting the Gaussian width of these radial scans as shown in Fig. \ref{BigPanel} (c), (f), (i).  Although triple-axis measurements do not have sufficient resolution to directly observe the peak splitting, the structural transition results in a significant change in extinction as shown in Fig. \ref{BigPanel} (b), (e), (h), (k), (m) where the temperature dependence of the (2,2,0)$_T$ peak intensity is plotted. Comparison of this data to the HB-3A measurements shows that the structural transition occurs at the inflection point of the temperature dependent intensity.

\begin{figure}
		\includegraphics[width=0.9\columnwidth]{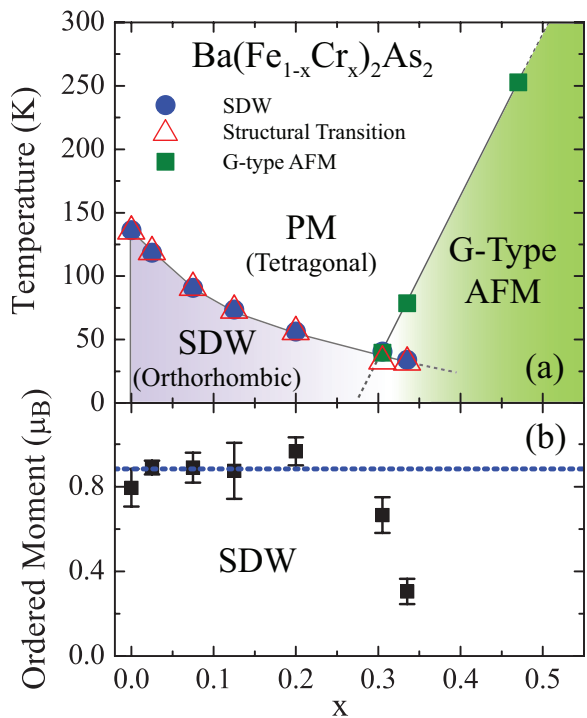}
		\caption{(Color online) (a) T-x Phase diagram of Ba(Fe$_{1-x}$Cr$_x$)$_2$As$_2$ indicating the SDW magnetic and concomitant structural phase transitions.  In addition, the phase boundary representing the transition from paramagnetic to G-type AFM order is shown. (b) The average ordered Fe/Cr moment as a function of x for the SDW state. The dotted line represents the average moment for x$\leq$0.2.}
		\label{PhaseDiagram}
\end{figure}

As can be clearly seen in Fig. \ref{BigPanel}, for 0$\leq$x$\leq$0.335 the SDW and structural transition always occur at the same temperature in these Cr-doped samples (with a maximum separation of $<$0.75 K for all samples with x$\leq$0.2).  This is in contrast to observations in the presence of Co, Ni, Rh, Pd, and Cu doping \cite{Fisher, Ni, Ni2} where the structural transition occurs at a higher temperature than the SDW.  To emphasize this point, Fig. \ref{BigPanel}(l),(m) shows measurements of the magnetic (1/2,1/2,3)$_T$ and nuclear (2,2,0)$_T$ reflections for a sample of BaFe$_{1.92}$Co$_{0.08}$As$_2$ measured in an identical manner demonstrating separated magnetic and structural transitions.  Similar observations of concomitant magnetic and structural transitions was recently reported for the case of Ru doping \cite{Thaler}. The presence of superconductivity in Ru-doped samples and its absence in samples doped with Cr indicates no correlation between the splitting of the transitions and superconductivity.  The resulting magnetic/structural transition is plotted as a function of concentration on the phase diagram of Fig. \ref{PhaseDiagram}(a).

To confirm the SDW magnetic structure and extract an ordered moment, several magnetic and structural reflections were measured above and below T$_{SDW}$.  The data were corrected for instrumental resolution using Reslib \cite{reslib}.  This confirmed that the arrangement of moments in the SDW state is identical to that of the parent compound for x$\leq$0.335.  The Fe/Cr ordered magnetic moment in the SDW state is extracted via normalization to nuclear reflections and plotted in Fig. \ref{PhaseDiagram}(b). The independent contributions from Fe and Cr cannot be distinguished due to the similarity of their magnetic form factors and, hence, the moment reported is the average for the Fe/Cr site.  Fig. \ref{PhaseDiagram}(b) shows a magnetic moment which is independent of concentration for x$\leq$0.2 despite suppression of the transition temperature from 140 K to 56 K over this concentration range.  This is in contrast to electron-doped BaFe$_2$As$_2$, \textit{e.g.}, in  Ba(Fe$_{1-x}$Co$_x$)$_2$As$_2$ the moment decreases monotonically with concentration and superconductivity only develops when the ordered moment has been reduced from that of the parent compound by a factor of $\sim$2.5\cite{Lester}. Cr-doping favors magnetism and stabilizes the moment over a large region of the phase diagram.  For x=0.305 and x=0.335, the ordered moment decreases rapidly with increasing concentration suggesting a destabilization of the SDW state.

\begin{figure}
		\includegraphics[width=1.0\columnwidth]{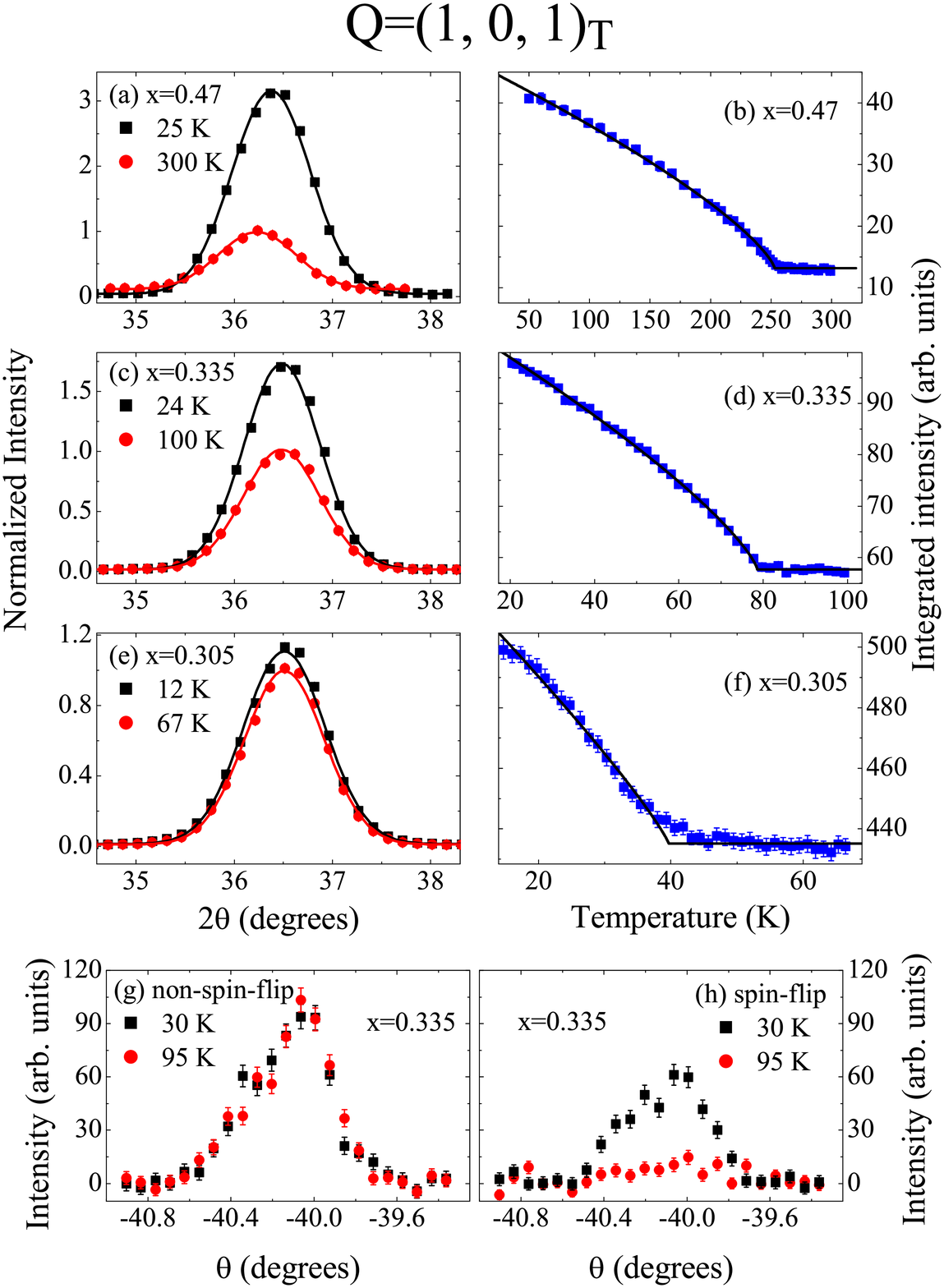}
		\caption{ (Color online) Measurements of the (1,0,1)$_T$ magnetic reflection representative of G-type magnetic order.  Radial scans for temperatures above and below T$_{N}$ are shown for (a) x=0.47, (c) x=0.335, and (e) x=0.305.  The temperature dependence of the integrated intensity of rocking curves through the (1,0,1)$_T$ reflection for (b) x=0.47, (d) x=0.335, and (f) x=0.305. Polarized neutron diffraction measurements of rocking curves through the (1,0,1)$_T$ reflection for a crystal with x=0.335 is shown in both (g) non-spin-flip and (h) spin-flip configurations for temperatures above and below T$_{N}$.  This shows the magnetic nature of the additional intensity.}
		\label{NNAFM}
\end{figure}

To understand the rapid decrease in moment with increasing concentration, supercell calculations for Ba(Fe$_{0.75}$Cr$_{0.25}$)$_2$As$_2$ were performed within the local spin density approximation, with fixed lattice parameters and relaxed atomic positions. These were done with the general potential linearized augmented plane wave method, as described in Ref. \cite{Sefat2}. Different magnetic orderings were considered (ferromagnetic, SDW and G-type AFM) and compared with the parent compound. Cr strongly favors a magnetic ground state.   In particular, the energy of antiferromagnetic solutions is much lower than paramagnetism, even more than in the pure Fe case, a consequence of the stronger covalency of Cr-As compared to Fe-As. These calculations predict that the two AFM solutions (SDW and G-type) are stable and are nearly degenerate at x=0.25. This competition between AFM states may provide an explanation for the rapid decrease in magnetic moment at higher Cr concentrations.

Motivated by these calculations, we performed measurements on crystals with x=0.2, 0.305, 0.335, and 0.47 mounted in the ($H$ 0 $L$)$_T$ scattering plane using the HB-3 TAS.  The intensity of the (1,0,1)$_T$ reflection is strongly enhanced on cooling, indicating a transition temperature which is higher (at least for x=0.335 and x=0.47) than that of the SDW state.  Furthermore, (1,0,0)$_T$ showed no significant temperature dependence suggesting a new magnetic ground state with a (1,0,1)$_T$ ordering wavevector.  As the $I4/mmm$ unit cell contains two atoms per Fe layer and two layers per unit cell, this ordering wavevector is consistent with G-type AFM order. Measurements of the temperature dependence for each of these concentrations is shown in Fig. \ref{NNAFM}(b),(d), and (f) indicating what appears to be continuous transitions into the G-type AFM state.  For both x=0.305 and x=0.335 this state coexists (either microscopically or as a separate phase) with the SDW state explaining the reduced SDW ordered moment in these two concentrations.

Unfortunately, magnetic reflections characterized by a (1,0,1)$_T$ propagation vector are coincident with nuclear reflections which, in some cases, are dominant.  However, the (1,0,1)$_T$ reflection is a strong magnetic peak and a relatively weak nuclear reflection allowing the magnetism to be easily measured.  This is not the case for any other reflections in the ($H$,0,$L$)$_T$ scattering plane making it difficult to establish definitely the magnetic nature of the extra intensity. Consequently, polarized neutron measurements were performed on the x=0.335 crystal.  The polarized results are summarized in Fig. \ref{NNAFM} (g) and (h)  where the non-spin-flip (g) and spin-flip (h) contributions are presented for temperatures above and below $T_N$ ($\sim$ 80 K for x=0.335).  These data show nuclear, non-spin-flip scattering which is unaffected by this transition while the magnetic, spin-flip scattering is strongly enhanced at low temperatures demonstrating the magnetic nature of the transition.

With only one clearly observed magnetic reflection, the magnetic moment direction and, hence, moment size cannot be uniquely determined in the G-type AFM state.  However, radial scans shown in Fig. \ref{NNAFM}(a), (c), and (e) give qualitative information on the strength of the magnetic intensity.  The high temperature, nuclear peak intensity has been normalized to unity in these panels showing magnetic intensity which is enhanced with increasing concentration and is almost 3 times the nuclear intensity for x=0.47.  This indicates either an ordered moment which increases as a function of concentration or an increased phase fraction for the G-type AFM order.

The observation of G-type AFM order in these three concentrations results in the phase diagram shown in Fig. \ref{PhaseDiagram}(a).  Measurements performed on the x=0.2 crystal indicate no evidence of the G-type AFM state for temperatures above 3.5 K.  This phase diagram confirms the DFT prediction that Cr-doping favors magnetism and that the SDW and G-type AFM ground state energies are comparable.  Undoped BaMn$_2$As$_2$ also exhibits G-type AFM order with a high transition temperature of 625 K \cite{bamn2as2,An}.  The steep slope of the paramagnetic to G-type AFM phase boundary is suggestive of a high transition temperature in the BaCr$_2$As$_2$ parent compound as well, which may indicate a similar phase diagram for the case of Mn-doping.

The results presented are summarized in the expanded phase diagram of Fig. \ref{PhaseDiagram}(a) revealing the presence of long-range magnetic order in Ba(Fe$_{1-x}$Cr$_x$)$_2$As$_2$ for all concentrations measured. The addition of Cr enhances magnetism, both of the SDW and competing orders, i.e. G-type.  In contrast to electron-doped materials, G-type magnetic order becomes the ground state at higher concentrations. These results provide compelling evidence that superconductivity is driven out by the stabilization of magnetism throughout the phase diagram.  This demonstrates the delicate balance between multiple competing states in the Fe-pnictides.

\begin{acknowledgments}
Research at ORNL is sponsored by the Scientific User Facilities Division and the Materials Sciences and Engineering Division, Office of Basic Energy Sciences, U.S. DOE. We acknowledge discussions with B.C Chakoumakos, B.C. Sales, M.A. McGuire, D. Mandrus, and V. Keppens and S. Kulan for technical support.  We thank J.W. Lynn for the use of his Heusler analyzer crystal.
\end{acknowledgments}


\end{document}